# Spin-torque switching in large size nano-magnet with perpendicular magnetic fields

Linqiang Luo[1], Mehdi Kabir[2], Nam Dao[3], Salinporn Kittiwatanakul[3], Michael Cyberey[4], Stuart A. Wolf[1, 3, 5], Mircea Stan[2] and Jiwei Lu[3]

[1]Department of Physics, University of Virginia, Charlottesville, Virginia, 22904, USA

[2]Department of Electrical & Computer Engineering, University of Virginia, Charlottesville, Virginia, 22904, USA

[3]Department of Materials Science & Engineering, University of Virginia, Charlottesville, Virginia, 22904, USA

[4]Department of Electrical Engineering, University of Virginia, Charlottesville, Virginia, 22904, USA

[5] Institute of Defense Analyses, Alexandria, Virginia, 22311, USA

Abstract:

DC current induced magnetization reversal and magnetization oscillation was observed in 500 nm large size $Co_{90}Fe_{10}/Cu/Ni_{80}Fe_{20}$ pillars. A perpendicular external field enhanced the coercive field separation between the reference layer ($Co_{90}Fe_{10}$) and free layer ($Ni_{80}Fe_{20}$) in the pseudo spin valve, allowing a large window of external magnetic field for exploring the free-layer reversal. The magnetization precession was manifested in terms of the multiple peaks on the differential resistance curves. Depending on the bias current and applied field, the regions of magnetic switching and magnetization precession on a dynamical stability diagram has been discussed in details. Micromagnetic simulations are shown to be in good agreement with experimental results and provide insight for synchronization of inhomogenieties in large sized device. The ability to manipulate spin-dynamics on large size devices could prove useful for increasing the output power of the spin-transfer nano-oscillators (STNOs).

Spin-polarized currents can be harnessed to manipulate magnetization and excite oscillation via the spin transfer torque (STT) effect, and are utilized in the application of MRAM[1,2] and spin-transfer nano-oscillators (STNOs)[3,4]. STNOs have the advantages that their frequencies are highly tunable by current and magnetic field over a range from a few GHz to 40 GHz.[3,5] Furthermore, the nanometer sized devices are among the smallest microwave oscillators yet developed[6] and their compatibility with standard silicon processing opens the possibility for on-chip applications.[7,8] However, the bottlenecks for the widespread application of STNOs lies in the enhancement of the output power above the current limit of ~ 0.5μW.[9] It has been suggested that two nano-contact STNOs in close proximity could mutually phase-lock and increase the output power; however phase-locking of more than two STNOs remains technologically challenge.[10-14] Instead of putting an array of STNOs nano-magnets together, we propose to make use of larger sized magnets in the hope that synchronization of multiple domains could lead to higher output power, and firstly we demonstrated that spin-transfer torque can be used to efficiently induce magnetization switching and oscillation in 500 nm large size devices. For large size device, our simulation results have shown that the non-uniform oscillations tend to synchronize with each other and generate coherent oscillation. In addition, large sized nano-magnets can be fabricated more cost-effectively through photolithography rather than using electron beam lithography.

The magnetic multilayer was synthesized by sputtering in a Biased Target Ion Beam Deposition system (BTIBD). The deposition details can be found elsewhere.[15] The complete structure of the multilayer is SiO$_2$ (substrate)/20nm Ru/2.2nm Co$_{90}$Fe$_{10}$ (reference layer)/5nm Cu/6.5nm Ni$_{80}$Fe$_{20}$ (free layer)/5nm Ru/Ti 5nm/Au 25nm. A magnetoresistance (MR) of ~1.2% was measured in the pseudo spin-valve continuous film using current-in-plane method before the patterning process [Fig. 1(a)]. The in-plane hysteresis loop revealed a good separation between the reference layer and free layer coercive fields of 30Oe and 5Oe, respectively. Then the magneto-transport behavior was characterized with a field applied in the out-of-plane (OOP) direction [Fig. 1(b)]. By applying an out-of-plane field, a much larger coercive field separation between the reference layer (~800 Oe) and free layer (~100 Oe) was achieved. The increase of the coercive fields can be understood in terms of the Stoner-Wohlfarth model.[16]

The film was then patterned using photolithography and then ion milled to form a round pillar with a nominal cross-section diameter of 500 nm. By monitoring the change of thickness and magnetization, we were able to precisely control the ion milling so that the $Co_{90}Fe_{10}$ layer was partially etched [as shown schematically in Fig. 1(c)]. With such geometry, the $Co_{90}Fe_{10}$ reference layer is relatively insensitive to the spin transfer torques due to its extended volume. Fig. 1(d) shows a scanning electron microscope image of the 500 nm-diameter size pillar. To reduce the resistance of the transmission lines, a very thick Au layer (4μm) was electro-deposited on the top contact areas, serving as the top bonding pads, as seen in the optical microscope image Fig. 1(e). The differential resistance is measured using a lock-in amplifier circuit in the PPMS system (PPMS 6000, Quantum Design). Positive current is defined in a direction such that electrons flow from $Co_{90}Fe_{10}$ to $Ni_{80}Fe_{20}$. All transport measurements reported in this work were obtained at room temperature. Micromagnetic simulations were carried out based on the LLG equation including a spin-torque term.[17] The parameters used to model the free layer were $M_S$=650 emu/cm$^3$, exchange stiffness constant of $A$=1.3×10$^{-6}$ erg/cm, and damping constant of $\alpha$ = 0.009. The mesh size was 10×10×3.25 nm$^3$. The spin polarization parameter $P$ was assumed to be 0.4.

The differential resistance $dV/dI$ versus perpendicular applied magnetic field $H$ is plotted in Fig. 1(f). The layers switch between antiparallel and parallel alignment with a current perpendicular to the plane (CPP) GMR ratio of 1.0%. The slightly lower GMR ratio after patterning might be due to the lead resistance that was in series with the pillar while it did not contribute to the MR. The coercive field of both layers increased significantly over the continuous film values. The coercive field of the $Co_{90}Fe_{10}$ reference layer was ~6.6kOe while the $Ni_{80}Fe_{20}$ free layer had a coercive field of ~1.3kOe. This is in accordance with observations on patterned nano-magnets, where the coercive field increases as it transitions from reversal through domain nucleation and domain wall motion in a full film towards coherent rotation in patterned structures.[18] As aforementioned, the reference layer had a significantly higher coercive field, allowing a large window of external magnetic fields for only the free-layer reversal. The gradual increase in resistance that occurs in Fig. 1(c) from 12kOe to 0Oe is from the anisotropic magnetoresistance (AMR) due to the remaining $Co_{90}Fe_{10}$ continuous film.

While both layers' magnetization tends to rotate toward the out-of-plane direction under the perpendicular field, the $Ni_{80}Fe_{20}$ layer's magnetization can be pulled more easily out of plane due to its smaller demagnetizing field ($4\pi M_S$). In comparison, a much larger external field was required to rotate the $Co_{90}Fe_{10}$ layer's magnetization. As a result, a pseudo hybrid configuration with an out-of-plane free layer and a near in-plane reference layer can be achieved at high perpendicular fields from ~3 kOe to ~9 kOe. The magnetic hybrid configuration has been predicted to be more efficient at generating large amplitude precession and therefore increase the output power in STNOs.[19-21]

Prior to the magneto-transport measurements, the sample was saturated under an out-of-plane field of 30 kOe, and then the field was set to the specific values. In Fig. 2(a), we show the variation of the differential resistance $dV/dI$ versus injected DC current at small fields. At $H = 0$, starting from the parallel (P) state at zero current, the DC current first decreased toward the negative direction. A jump in the differential resistance was observed at $I_C^{P-AP}$ = -48mA (current density of $2.45\times10^7 A/cm^2$) due to the magnetization switching which occurred in the $Ni_{80}Fe_{20}$ layer. The curve showed hysteretic behavior because the system remained in the antiparallel (AP) state until the current was swept back to a positive critical current of $I_C^{AP-P}$ = 47mA (current density of $2.39\times10^7 A/cm^2$), where the resistance dropped back to the P state level. The corresponding change in the differential resistance was ~ 95mΩ for both negative and positive switching, which was of the same magnitude of change in resistance shown in the MR minor loop [Fig. 2(a) inset]. Therefore, it was certain that the experimental hysteretic curve was caused by spin torque induced magnetization reversal typically observed at zero or low applied magnetic field.[3] The hysteretic reversal here was fundamentally different from the switching by the Oersted field that was self-generated by the current passing through the pillar.[22] The magnitude of the Oersted field was independent of the current direction, which would have resulted in a symmetric *dV/dI* curve with respect to current. Note that the overall parabolic increase in *dV/dI* can be ascribed to the electron scattering by emissions of phonons and magnons in metallic point contacts.[23]

Micromagnetic simulations accurately reproduced the dynamic process of the current induced switching observed experimentally. Fig. 2(b) shows a typical switching from AP to P reproducing the behavior shown in the Fig. 2(a) blue curve. For our device, the Oersted field is not strong enough to form a full vortex at the experimental currents used (~48 mA). The micromagnetic simulation suggests a formation of a full vortex by the Oersted field requires a current of at least ~100 mA, which agrees with a previous report.[24] Instead, a *C*-shape magnetization state tends to form under the Oersted fields, as shown in Fig. 2(b). The *C*-state has a majority magnetization pointing along the parallel or anti-parallel direction. Under the action of the spin-transfer torque, the *C*-state rotates toward the opposite direction leading to the switching from antiparallel to parallel.

Starting from a parallel alignment of the magnetic layers, a series of *dV/dI* versus $I_B$ scans were measured at different fields to construct the $I_B$-*H* dynamic stability diagram shown in Fig.3(a). Each symbol in the diagram corresponds to a discrete change in the resistance while changing $I_B$ at a corresponding field. We discuss the results in Fig. 3 in details below; the general features of the dynamic stability diagram are also seen in 130 nm × 70 nm Co/Cu/Co devices.[3] Under 3.2 kOe, we observe hysteretic reversal of magnetization between the AP and P states where the red circles correspond to the critical current for AP-P transition $I_C^{AP\text{-}P}$ and blue squares correspond to the critical current for P-AP transition $I_C^{P\text{-}AP}$. For *H* > 3.2 kOe we observe magnetic transitions that are reversible in $I_B$ scans and give sharp peaks in $dV/dI$. A reversible change is demonstrated in Fig. 3(b) with the $dV/dI$ curve recorded at $H = 3.5$ kOe.

For $H < 3.2$ kOe, the change of critical current $I_C^{AP-P}$ follows a linear $H$ dependence. This agrees with the critical currents equations given by the Slonczewski model:[25,26]

$$I_C \approx \frac{A\alpha M_s V}{g(\theta)P}(-H_K + 4\pi M_s - H_{dip} - H) \qquad (1)$$

where $M_s$, $V$ and $\alpha$ are the saturation magnetization, volume and Gilbert damping constant for the free layer, respectively. $A$ is a constant coefficient of the order of $10^{11}$ mA Oe$^{-1}$ emu$^{-1}$, $P$ is the spin polarization, $g(\theta)$ is a scalar factor depending on the relative angle of the reference layer and free layer magnetizations.[26] $H$, $H_{dip}$ and $H_k$ are the perpendicular field, the dipolar field and the anisotropy field, respectively, while $4\pi M_s$ arises from the demagnetizing field. $I_C^{AP-P}$ follows the linear $H$ dependence with a slope = $-3.25 \times 10^{-2}$ mA/Oe and intercepts $I_B = 0$ near 2 kOe, which is roughly the positive coercive field as shown in Fig. 2(a) inset. By extracting the slope of $I_C^{AP-P}$ and comparing with the pre-factor in Eq. (1), we estimate $\alpha/P = 0.42g(\pi)$. Given that $\alpha = 0.009$ and $P = 0.4$,[27] then $g(\pi)=0.054$, which is almost an order smaller than the expected range between 0.1 and 0.5.[28] The difference might be associated with the multiple domains in our large size device. According to the Slonczewski model, Eq. (1) is strictly valid for single domain structure at low temperatures.[26] Hence, Eq. (1) may not be directly applicable to the possible multi-domain structures in a 500 nm magnetic nanopillar.

For $H > 3.2$ kOe, the hysteretic switching of the differential resistance is replaced by the peaks given by the green triangle symbols in the dynamic stability diagram. It is found that these peaks are generated only on the negative currents direction which corresponds to the electrons flow direction from $Ni_{80}Fe_{20}$ to $Co_{90}Fe_{10}$. This feature is also a signature of the spin transfer torque, different from the effects of the Oersted fields.[29] The change from irreversible hysteretic switching to reversible sweeping of the differential curve has been associated in standard nanopillars to the sustained precession of magnetization.[3,5]

In the dynamic stability diagram, the $I_C^{P\text{-}AP}$ and $I_C^{P\text{-}AP}$ curves are extrapolated to construct the oscillation regime [the blue regime shown in Fig. 3(a)]. The peaks generally shift to higher current with increasing fields following the trend of the extrapolated curve.

Examining Fig. 3(b), we observe multiple peaks at 3.5 kOe. Furthermore, these peaks have different critical currents and show distinct amplitudes as shown in the second derivative curve Fig. 3(b) inset. Given the large size of our device, inhomogeneities, such as multiple domains, exist on the ferromagnetic layers. However, the micromagnetic simulations showed that the multiple domains tended to synchronize with each other under certain currents and led to a harmonic precession as demonstrated in Fig. 4(a). A non-coherent multiple domain precession occurred at the beginning Fig. 4(b) and evolved into a coherent oscillation as shown in Fig. 4(c). Based on the simulation, the multiple peaks are likely to be associated with different spin-wave modes that are represented by the different precession orbits as shown in Fig. 4(c). The orbits shown are numerically calculated from the LLGS equation with out-of-plane field of 3.5kOe. A large external field could suppress the number of domains and possibly lead to the single mode precession as shown in the dynamic stability diagram for $H = 4.5$ kOe and 5 kOe, yet, more work needed to be carried out to confirm the spin dynamics experimentally on the frequency domain.[30,31]

In summary, 500 nm large size $Co_{90}Fe_{10}/Cu/Ni_{80}Fe_{20}$ pseudo spin valve pillars were fabricated using only photolithographic techniques. We demonstrate the magnetization reversal and magnetization oscillation on these large sized devices. At large fields, only the negative polarity currents cause peaks in the *dV/dI*, which is consistent with the excitation of precessing spin wave modes. The critical current for the magnetization switching and oscillation have been plotted together to construct the dynamic stability diagram. In contrast to the dynamic diagrams of standard nanopillars, the switching regime of our devices show deviation from the Slonsczewski model, which might be due to the multiple domains of our large size pillar. Our simulations indicate that the multiple peaks could come from the different spin-wave modes with independent precession orbits. The spin torque oscillation in large size devices could lead to potential applications that require enhanced power of the STNOs.

**Acknowledgement:**
Authors are grateful to the support from the National Science Foundation (Award No. ECCS-1344218). J.L. thanks the support from C-SPIN center.

**Figure captions:**

Fig 1. (a) & (b) Current in-plane (CIP) magnetoresistance (MR) curve on the full film before patterning with in-plane field or out-of-plane field respectively. (c) Schematic representation of the patterned pseudo-spin-valve sample. The bottom $Co_{90}Fe_{10}$ layer is partially patterned. The arrow shows the electron flow direction for negative current. (d) A SEM image of the patterned pillar with a diameter of 500 nm. (e) The optical microscope image of a device with CPW contacts. (f) Current perpendicular to plane (CPP) MR curve on the patterned sample. The black arrows show the direction of the field sweep.

Fig 2. (a) Differential resistance versus biased direct current ($I_B$) with out-of-plane magnetic fields of 0 (bottom, red) and 200 Oe (blue). The current sweep starts at 0 mA and the arrows indicate the direction of the current sweep. The inset shows the minor MR hysteresis loop for $Ni_{80}Fe_{20}$ layer. The black arrows indicate the direction of the field sweep. (b) Micromagnetic simulation showing the evolution of $m_x$ component at a current density of $3 \times 10^7 A/cm^2$ with an out-of-plane field $H = 200$ Oe. The flip-over of a *C*-state is demonstrated by the spatial magnetization distribution 1→2→3→4. The cores are added for illustration purpose.

Fig 3. (a) Experimentally determined $I_B - H$ (Out-of-Plane field) dynamic stability diagram for the 500 nm pillar device where each symbol corresponds to a distinct change in the resistance. Below 3.2 kOe, the symbols are from hysteretic switching where the circles (red) are $I_C^{AP-P}$ and squares (blue) are $I_C^{P-AP}$. For field larger than 3.2 kOe, the triangles (green) refer to the peaks on the differential resistance curves. (b) *dV/dI* versus $I_B$ for $H = 3.5$kOe corresponding to the field position marked by the dashed line in (a). The inset shows the second derivative of *dV/dI* indicating the relative amplitude, width and position of the peaks.

Fig 4. (Color online) (a) Micromagnetic simulation showing the temporal magnetization component evolutions of $m_x$ (red), $m_y$ (blue) under an out-of-plane field of 3.5 kOe and with a constant current density of $3.0 \times 10^7 A/cm^2$. (b)&(c) The evolution of the spatial magnetization distribution at the corresponding time marked under them. (d) At $H = 3.5$ kOe, different current density leads to distinct stabilized precession orbits and frequencies. The main precession frequency $f$ is determined by taking the Fourier transforms of $m_x$. At $2.0 \times 10^7 A/cm^2$ (black), $3.0 \times 10^7 A/cm^2$ (red) and $3.4 \times 10^7 A/cm^2$ (blue), the simulated oscillation has the frequency of 3.15GHz, 10.5GHz and 14.6 GHz respectively.

*Fig. 1*

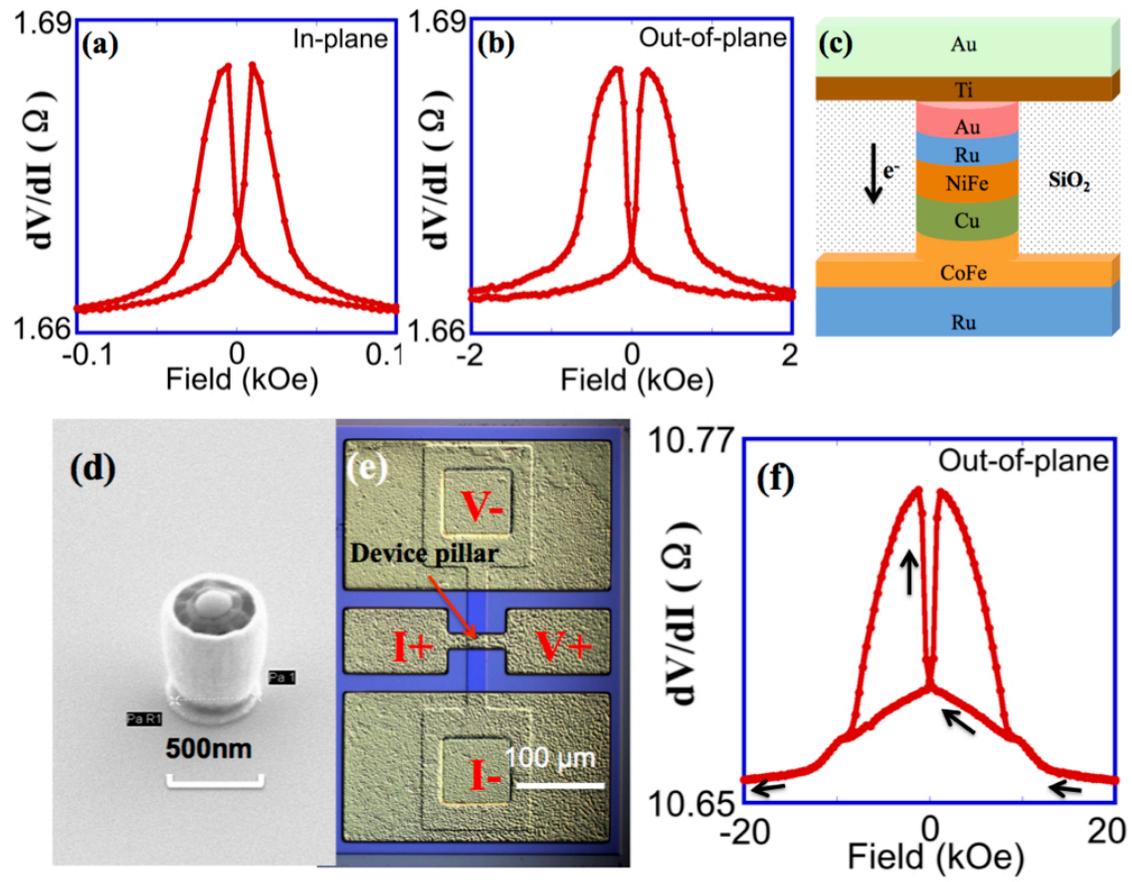

*Fig. 2*

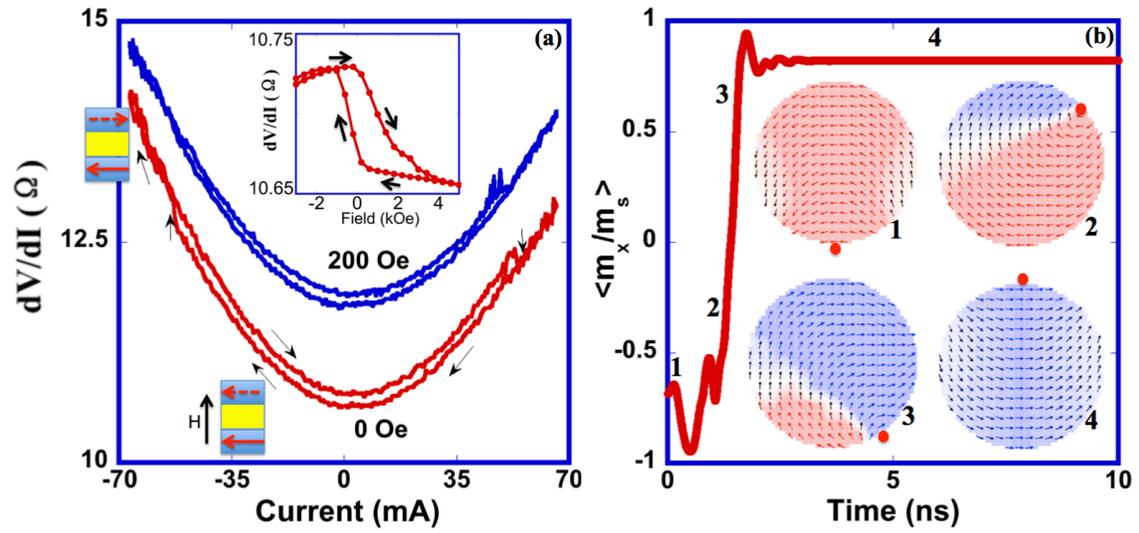

*Fig. 3*

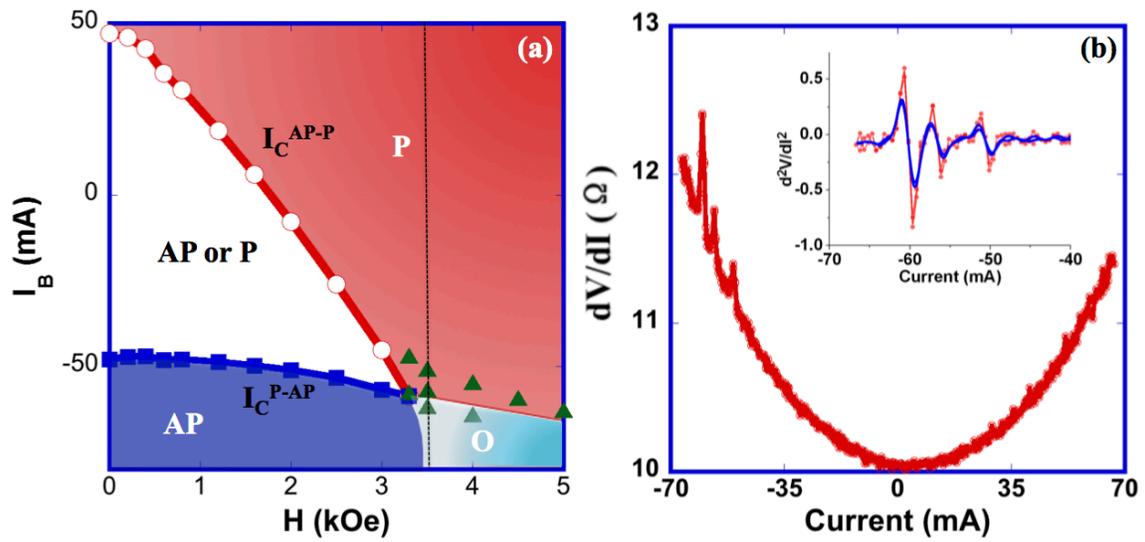

*Fig. 4*

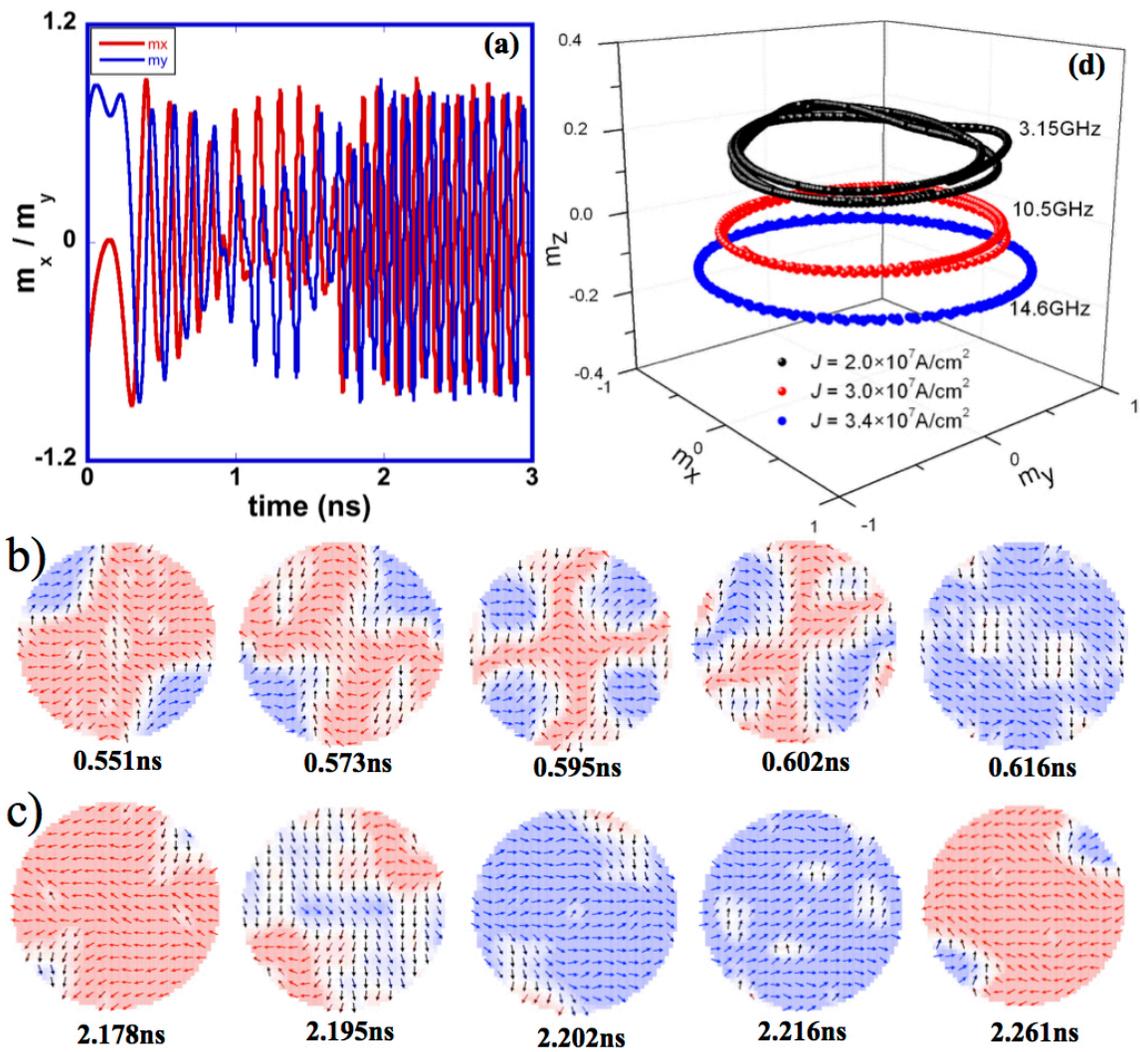